\documentclass[10pt,notitlepage, superscriptaddress]{revtex4-1}

\usepackage[latin2]{inputenc}
\usepackage{amsmath}
\usepackage{amssymb}
\usepackage{bbm}
\usepackage{graphicx}
\usepackage{color}

\graphicspath{{figures/}}
\usepackage{subfigure}
\newlength{\figurewidth}
\newcommand{\Tr}{\mathrm{Tr}}
\newcommand{\ket}[1]{\left|#1\right\rangle  }
\newcommand{\bra}[1]{\left\langle#1\right|  }
\newcommand{\ro}{\varrho}
\newcommand{\roh}{{\tilde\varrho}}
\newcommand{\Mc}{\mathcal{M}}
\newcommand{\Ml}{M_\lambda}
\newcommand{\Mlk}{M^\dag_\lambda}
\newcommand{\R}[2]{{(\roh^{t}_{#1}+\roh^{t}_{#2})}}
\newcommand{\betac}{\bar{\beta}}
\newcommand{\Deltac}{\bar{\Delta}}

\usepackage{hyperref}

\begin{document}

\title{Non-Markovian Toy Quantum Chain}

\date{\today}

\author{András Bodor}
\affiliation{
Department of Physics of Complex Systems, E\"otv\"os University\\
H-1117 Budapest, P\'azm\'any P\'eter s\'et\'any 1/A, Hungary}
\author{Lajos Di\'osi}
\affiliation{
Research Institute for Particle and Nuclear Physics\\
H-1525 Budapest 114, POB 49, Hungary}
\author{Zs\'ofia Kallus}
\affiliation{
Department of Physics of Complex Systems, E\"otv\"os University\\
H-1117 Budapest, P\'azm\'any P\'eter s\'et\'any 1/A, Hungary}
\author{Thomas Konrad}
\affiliation{
School of Chemistry and Physics, University of KwaZulu-Natal\\
Private Bag X54001, Durban 4000, South Africa}

\begin{abstract}
We propose a simple structure for stationary non-Markovian quantum chains in the
framework of collisional dynamics of open quantum systems. To this end, we modify the
microscopic Markovian \mbox{system--reservoir} model, consider multiple collisions
with each of the molecules with an overlap between the collisional time intervals. We
show how the equivalent Markovian quantum chain can be constructed with the addition
of satellite quantum memory to the system. We distinguish quantum from classical 
non-Markovianity. Moreover, we define the counts of non-Markovianity by the required number of 
satellite qubits and bits, respectively. As the particular measure of quantum 
non-Markovianity the discord of the satellite w.r.t. the system is suggested.
Simplest qubit realizations are discussed, and the significance for real 
\mbox{system--environment} dynamics is also pointed out.
\end{abstract}
\maketitle

\section{Introduction}\label{Introduction}
In recent years we have seen increasing interest in abstract 
modeling of time evolution of open quantum systems coupled to
their environment via different quantum channels. 
While decrypting fundamental laws is their ultimate goal, 
inspirations of such models and possible future applications of their results range 
from solid state physics to quantum biology and quantum information technology
\cite{Weixx,BreuerPetruccione,WisMil09,RivHue11}.

Dynamics modeled microscopically as a series of discrete interactions or ``collisions''
between a central system and environmental (reservoir) \emph{molecules}
leads to abstract collisional system-reservoir models. For a memoryless 
Markovian time evolution the reservoir has the idealized ability of 
\emph{forgetting} new information faster than the rate of collisions;
i.e., after each collision the reservoir would totally relax to its pre-collisional state 
before the next collision occurs.
For the (central) system, this results in what we call Markovian quantum chain.
In each step along the chain, the system undergoes decoherence, 
leakage of information out to the reservoir to be forgotten there.  
Mathematically, the Markov quantum
chain's irreversible dynamics is obtained from reduction of the unitary time evolution
of the system-environment compound by means of partial trace over degrees of freedom of the environment.
In the special case of stationarity, the mathematical tool of semigroups of completely
positive (CP) trace preserving dynamical maps can be used
\cite{BreuerPetruccione,WisMil09,RivHue11,NieChu00,Dio11}.

The collisional model of Markovian chains has been studied in numerous works \cite{collisionalModels,collisionalModelsDio}. 
It will be particularly useful if we want
to monitor the system \cite{Dio02,AudDioKon02,Dio03}. Although we are never allowed to directly measure the system state,
we can measure each reservoir molecule after its collision with the system. 
By their regular selective measurements we can monitor the system state indirectly. 
The corresponding resolved evolution of the quantum chain is called 
selective Markovian quantum chain.

A more realistic description requires an account of the reservoir memory and the induced 
non-Markovianity of the dynamics of the central system (the chain). 
For analyzing new dynamical properties and quantifying 
the difference from the simple Markovian processes, 
several non-Markovianity measures have been proposed
so far. 
These make use of different aspects of non-Markovian (NM)
evolutions, e.g., the non-divisibility of the underlying 
quantum dynamical map \cite{Wolf08}, the increasing trace distance 
(i.e.: distinguishability) of two initial quantum states, accompanied by
the back-flow of information from the reservoir to the system
\cite{Breu0910}, or the \emph{discord} between system and reservoir
\cite{Alipour12_NMdiscord}.
Non-Markovianity is a field of active debates (cf., e.g.,
\cite{Breuer11_NMmeas}) so we emphasize the importance of abstract 
models to capture fundamental aspects of memory keeping processes.

\begin{figure*}[tbp]
\begin{center}
\subfigure[]{
    \includegraphics[width=0.8\figurewidth, angle=0]{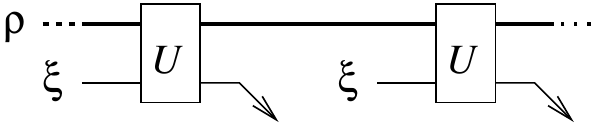}
    \label{fa_markov}}
\quad
\subfigure[]{
  \includegraphics[width=0.8\figurewidth, angle=0]{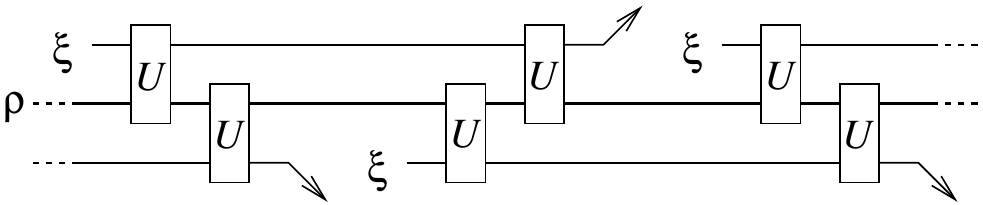}
  \label{fa_diosi}}
\quad
\subfigure[]{
  \includegraphics[width=0.8\figurewidth, angle=0]{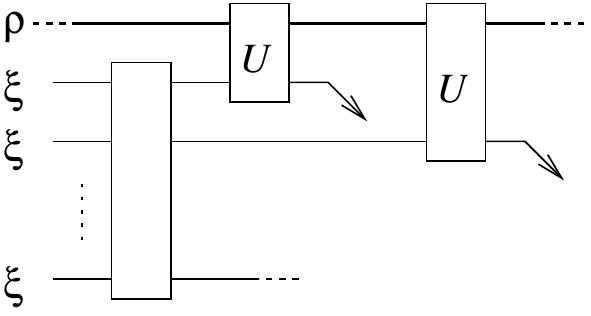}
  \label{fa_buzek}}
\quad
\subfigure[]{
  \raisebox{3em}{\includegraphics[width=0.8\figurewidth, angle=0]{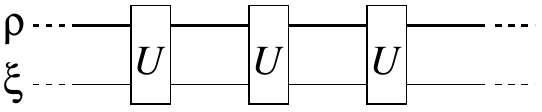}}
  \label{fa_chiuri}}
\caption{ 
Quantum chains: different system-reservoir collision models
($\ro$: system state, $\xi$: reservoir molecule state, 
$U$ system-molecule unitary collision operator).   
\subref{fa_markov} Standard Markovian chain: 
the system interacts with each independent molecule once.
\subref{fa_diosi} The proposed NM chain: the system
interacts with each independent molecule twice, with overlap
between collisional intervals of 'nearest' molecules. 
\subref{fa_buzek}  NM chain  in \cite{Ryb12}: the system
interacts with each molecule only once, but the molecules
are initially entangled.
\subref{fa_chiuri} NM chain in \cite{Chi12}:  
the system interacts with the same molecule multiple times.
}
\label{f_familyalbum}
\end{center}
\end{figure*}

In this paper we study the structural features of NM open quantum systems 
by constructing discretized NM processes, i.e., quantum chains. 
Starting from the generic structure of a Markovian quantum chain 
we are going to impose a certain NM structure on it. The proposed 
abstract NM structure corresponds to an open quantum system in a 
reservoir of non-interacting molecules which only collide with 
the central system through unitary collisions. The initial, for the 
moment Markovian, quantum chain can be seen in Fig.~\ref{fa_markov}.
To engineer the NM element, we allow each molecule to
collide with the system more than a single time, say twice, while the
molecules' collisional intervals overlap, by assumption, with
the collisional periods of both previous and consecutive molecules, 
cf. Fig.~\ref{fa_diosi}. 
Instead of separating what is forgotten and what is not by different
timescales, our abstract model has a built-in exact memory time.
Most importantly, we show how our NM chains become Markovian at the price
of attaching \emph{satellite memory} subsystems. Due to our proposed specific
NM structure, the identification of the satellite memory within the environment
is straightforward compared especially to the efforts requested in oscillatory
reservoir models (cf. \cite{Wooetal11,Dio12} and Refs. therein).
Then a natural \emph{count} of non-Markovianity follows: let it be the size 
(in qubits) of the requested satellite memory. We can refine this count
into informatic \emph{measures} of non-Markovianity, as we show later. 

The NM quantum collisional dynamics is not new in itself.
Alternative NM structures are shown in Figs.~\ref{fa_buzek} and \ref{fa_chiuri}. 
Ref.~\cite{Ryb12} introduces NM mechanism by starting all molecules from an 
a priori entangled state, cf. Fig.~\ref{fa_buzek}. This work focuses on the molecular
realization of a single NM step instead of general NM chains. In \cite{Chi12}, 
the environment consists of a single molecule and this molecule collides with 
the system consecutively, many times, 
cf. Fig.~\ref{fa_chiuri}. An important advantage of our NM structure over 
these two is that it can invariably host the monitoring of the system, 
which was so instrumental for quantum Markov chains but becomes complicated 
in structure Fig.~\ref{fa_buzek} and impossible in Fig.~\ref{fa_chiuri}. 

In Sec.~\ref{Quantum_Markov_Chain} the exact notion of a non-selective and
a selective quantum Markovian chain is defined. The definition is extended for 
quantum NM chains in Sec.~\ref{Quantum_non-Markov_Chain}. 
Then Sec.~\ref{One-qubit_non-Markov_Chains} contains our new results demonstrated
on the toy models of one-qubit NM chains. Based on these results  we discuss in 
Sec.~\ref{Classical_Quantum_non-Markovianity} a new measure of quantum non-Markovianity 
as well as a distinction between classical and truly quantum non-Markovianity.

\section{Quantum Markov chain}\label{Quantum_Markov_Chain}
To define Markov chains, we apply the toolbox containing CP-maps,
Kraus matrices, POVMs, selective and non-selective quantum measurements,
detailed by the monographs \cite{BreuerPetruccione,WisMil09,RivHue11,NieChu00,Dio11}.
We understand by a non-selective Markov chain 
a series of quantum states $\ro_0,\ro_1,\dots,\ro_t\dots$ of a given system
where each state along the chain depends but on the preceeding state respectively:    
\begin{equation}\label{MC}
\ro_{t+1}=\Mc_t\ro_t;~~~~~~t=0,1,2,\dots,
\end{equation}
where $\Mc_t$ are CP maps.
If not stated otherwise, we restrict ourselves to stationary chains $\Mc_t=\Mc$.
A CP map can always be represented by certain Kraus
matrices $\Ml$ such that 
\begin{equation}\label{Kraus}
\Mc\ro=\sum_\lambda\Ml\ro\Mlk .
\end{equation}
Accordingly, non-selective Markov chains (\ref{MC}) can be decomposed 
into selective Markov chains given by the recursion relation:
\begin{equation}\label{SMC}
\ro_{t+1}=\frac{1}{p_{\lambda,t}} \Ml \ro_t \Mlk .
\end{equation}
$\lambda$ may represent the random outcome of a POVM measurement
characterized by the matrices $\Ml$. Here
$\Tr\Ml\ro_t\Mlk$ is the outcome probability. 
As a matter of fact, 
the selective chain $\ro_t$ depends on all measurement outcomes $\lambda$
prior to $t$, this dependence is suppressed in our notation.
The stochastic average of the selective Markov chain (\ref{SMC}) over the
measurement outcomes $\lambda$ yields the non-selective Markov
chain (\ref{MC}).

We can always construct a microscopic model for a given quantum Markov chain. Consider an abstract ideal gas (reservoir) of 
identical molecules each in state $\xi$. 
Independent unitary collisions will generate the CP map $\Mc$ of (\ref{MC}):  
\begin{equation}\label{trUMC}
\ro_{t+1}=\Tr_{\mathrm{res}}[U(\ro_t\otimes\xi)U^\dag] ,
\end{equation}
where $U$ is the collision matrix, $\Tr_{\mathrm{res}}$ is the partial trace over the molecule state. 
The process is shown in Fig.~\ref{fa_markov}.
Any map (\ref{Kraus}) can be generated 
by a suitable unitary mechanism (\ref{trUMC}) whereas the choice of $U$ and $\xi$
is never unique \cite{NieChu00,Dio11}.
If we inspect, i.e. measure, the post-collisional state of 
the molecule we get the underlying microscopic model of the selective
Markov chain (\ref{SMC}), 
\begin{figure}
\centering
\includegraphics[scale=1.0]{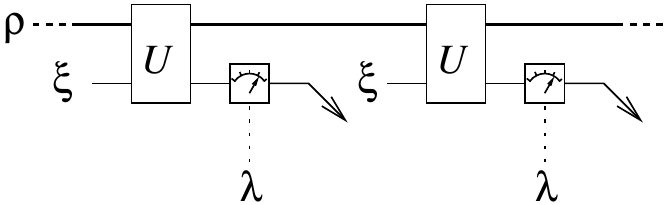}
\caption{Selective Markov chain: the refinement of the non-selective one in Fig.~\ref{fa_markov}.
The post-collisional state of the molecule is measured, and yields the random outcome $\lambda$.}
\label{f_SMC}
\end{figure}
shown in Fig.~\ref{f_SMC}. As to determining Kraus matrices $\Ml$, their possible choice is
simple if the molecules are prepared in pure state $\xi=\ket{\Psi}\bra{\Psi}$:
\begin{equation}\label{MlUPsi}
\Ml=\bra{\lambda}U\ket{\Psi}.
\end{equation}
We show concrete examples in Sec.~\ref{One-qubit_non-Markov_Chains}.

\section{Quantum non-Markov chain}\label{Quantum_non-Markov_Chain}
Consider the chain
\begin{equation}\label{NMC}
\ro_t=\Mc(t)\ro_0;\quad t=0,1,2,\dots,
\end{equation}
where $\Mc(t)$ is a $t$-dependent CP map.
The chain is Markovian, see (\ref{MC}), if $\Mc(t)$ is \emph{divisible}, i.e. can be written in the following form:
\begin{equation}\label{MCdef}
\Mc(t)=\Mc_t\Mc(t-1);~~~~~~t=1,2,\dots,
\end{equation} 
with some sequence of CP maps $\Mc_1,\Mc_2,\dots,\Mc_t,\dots$. 
To be clear, we require that
\begin{align}
\nonumber\Mc(1)&=\Mc_1, \\
\Mc(2)&=\Mc_2\Mc_1, \\
\nonumber\Mc(3)&=\Mc_3\Mc_2\Mc_1,~~~\mathrm{e.t.c.}
\end{align}
If such a sequence $\Mc_1,\Mc_2,\dots,\Mc_t,\dots$
does not exist, i.e., when $\Mc(t)$ is not divisible into the same set of factors for all $t$, 
the chain $\ro_t$ is NM.

The microscopic mechanism of quantum non-Markovianity can be quite complicated.
A recent study \cite{Ryb12} considers single qubit maps which are not divisible at all. 
To model such maps microscopically, a class of finite NM chains
has been constructed. Its key mechanism has been independent collisions 
with entangled molecules, the general structure is shown in Fig.~\ref{fa_buzek}.
Another work \cite{Chi12} considered a single qubit environment, interacting unlimited
times with the system qubit via a controlled-rotation.
Our approach will be different and elementary. 
We construct a class of stationary non-Markovianity.
Our molecules are independent, and do not become entangled prior to their collisions with the
system. However, they will collide multiple times --- twice, for simplicity's sake --- and this
constitutes the memory mechanism. 
Fig.~\ref{fa_diosi} shows the general NM structure. 
\begin{figure}
\centering
\includegraphics[scale=1.0]{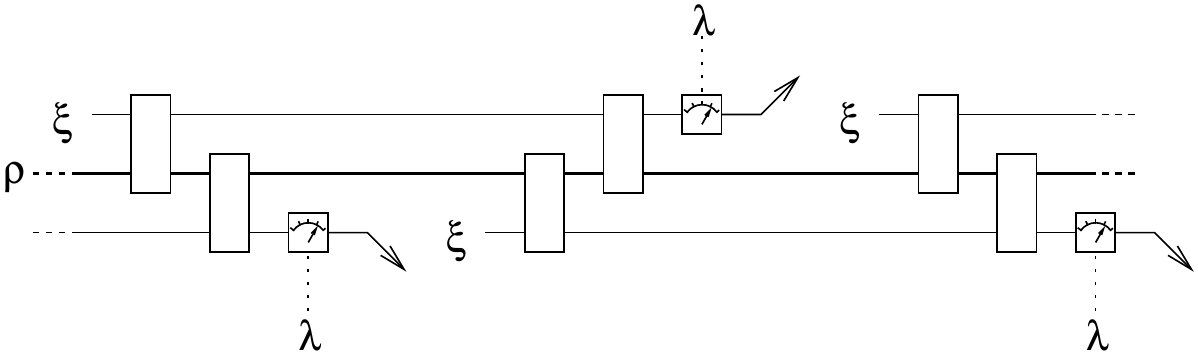}
\caption{Selective NM chain: the refinement of the non-selective one in Fig.~\ref{fa_diosi}.
The post-collisional (i.e.: after the double collision) state of the molecule is measured, 
and yields the random outcome $\lambda$.}
\label{f_SNMC}
\end{figure}
Unlike the previous NM structure shown in Fig.~\ref{fa_chiuri}, 
ours allows us, by its construction, to insert measurements without
changing the physics of the process. Obviously, we can insert measurements
on each molecule after its second collision, as shown in Fig.~\ref{f_SNMC}.

Rather than discussing this NM model generally, in what follows we concentrate on 
its simplest qubit realizations.

\section{One-qubit non-Markov Chains}\label{One-qubit_non-Markov_Chains}
In order to highlight the similarities and differences, first, in Sec.~\ref{One-qubit_Markov_chain}, 
we are going to construct a one-qubit Markov chain, and then, 
in Secs.~\ref{Repeated_XORs} and \ref{Distributed_XORs}, by extending it, we study two different NM structures.

Throughout this section the molecules constituting the reservoir are single qubits in pure initial states
$\xi=\ket{\Psi_{\phi}}\bra{\Psi_{\phi}}$, where 
\begin{equation}\label{ketphi}
\ket{\Psi_{\phi}}=\cos\phi\ket{0} + \sin\phi\ket{1} .
\end{equation}
For later convenience, we write the state of the molecule as 
\begin{equation}\label{phiket0}
\ket{\Psi_{\phi}}=\exp(i\phi\sigma_y)\ket{0}.
\end{equation}

\subsection{One-qubit Markov chain}\label{One-qubit_Markov_chain}
Let our central system be a single qubit, and let us construct a Markov chain (\ref{MC}-\ref{SMC}), 
also see Figs.~\ref{fa_markov} and \ref{f_SMC}.
Couple the molecule in state $\xi$ to the system qubit in state $\ro_t$
via the XOR-gate.
\begin{figure}
\centering
\includegraphics[scale=1.0]{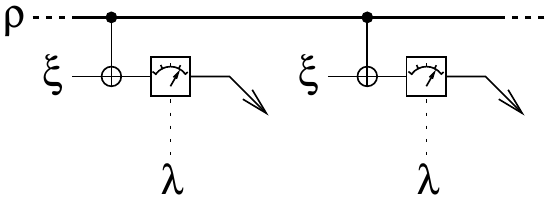}
\caption{Selective one-qubit Markov chain: collisional model. The system qubit $\ro$
interacts (collides) with the reservoir qubits (molecules) $\xi$ via XOR-gates; 
the post-interaction states of the molecular qubits are measured, yielding the random
outcomes $\lambda$.}
\label{f_SMC_qub}
\end{figure}
The corresponding two-qubit unitary operator $U$ is the following:
\begin{equation}\label{U2XOR_1}
U=\mathrm{XOR}_\mathrm{sys-mol}.
\end{equation}
The XOR-gate installs the unsharp measurement of the system qubit by the molecular
qubit. The strength of the measurement is controlled by the parameter $\phi$ of
the molecular pre-collisional state. If $\phi=0,\pi/2$, the measurement is projective
while $\phi\approx\pi/4$ yields a weak measurement \cite{BruDioStr08}; the value $\phi=\pi/4$ decouples
the system from the molecules. 
 
The two Kraus matrices are defined by $\Ml=\bra{\lambda}U\ket{\Psi_\phi}$ for
$\lambda=0,1$, cf. Eq.~(\ref{MlUPsi}). Inserting (\ref{phiket0}), we have
\begin{equation}\label{Ml_def}
\Ml=\bra{\lambda}U\exp(i\phi\sigma_y)\ket{0} .
\end{equation}
To perform the partial trace operation
(\ref{trUMC}) from the four-dimensional state $U(\ro_t\otimes\xi)U^\dag$ to the two-dimensional 
state $\ro_t$, we use the deeper level selective form (\ref{SMC}) for technical reasons. 

The unitary evolution of the molecule-system compound $\xi\otimes\ro_t$ in the
product basis $\ket{\mathrm{mol,sys}}=\{ \ket{00},\ket{01},\ket{10},\ket{11}\}$ is given by
\begin{equation}\label{UXOR}
U\exp(i\phi\sigma_y)=
\begin{pmatrix} 
  \cos\phi & & -\sin\phi & \\ 
   & \sin\phi & & \cos\phi \\
   \sin\phi & & \cos\phi & \\ 
   & \cos\phi & & -\sin\phi \\ 
\end{pmatrix} .
\end{equation}
The two Kraus matrices in this representation are simply
the left block matrices of $U$:
\begin{equation}\label{M2x2}
M_0=
\begin{pmatrix} 
  \cos\phi &  \\ 
           & \sin\phi
\end{pmatrix}
\hspace{3em}
M_1=
\begin{pmatrix} 
  \sin\phi &  \\ 
           & \cos\phi
\end{pmatrix} .
\end{equation}
The non-selective
Markov chain (\ref{MC}) is as trivial as \footnote{Be cautious, we swing from
notation $\ro_t$ to $\ro^t$ whenever we must indicate matrix indices as well.}
\begin{equation}\label{MCcomp}
\ro^{t+1}_{00}=\ro^{t}_{00} \hspace{1em}  \ro^{t+1}_{11}=\ro^{t}_{11} \hspace{1em} \ro^{t+1}_{01}=\sin(2\phi) \ro^{t}_{01}
\end{equation}
i.e., the diagonal elements are preserved while the off-diagonals will
step towards zero unless we took the singular molecular states with 
$\phi=\pi/4$:
\begin{equation}\label{rodiag}
\ro_\infty=
\begin{pmatrix}
\ro^0_{00}          & 0\\
 0 & \ro^0_{11}
\end{pmatrix} .
\end{equation}
We can say that our Markov chain with $\phi\neq\pi/4$ is 
asymptotically equivalent with a single von Neumann projective measurement.
Even if the single collisions are in the weak measurement regime, their
cumulative effect is the projective measurement, as is well known, e.g., 
from \cite{Dio02,AudDioKon02,Dio03}.

\subsection{Repeated XORs with time overlap}\label{Repeated_XORs}
\begin{figure*}[tbp]
\begin{center}
\subfigure[NM qubit chain]{
    \includegraphics[scale=1.0]{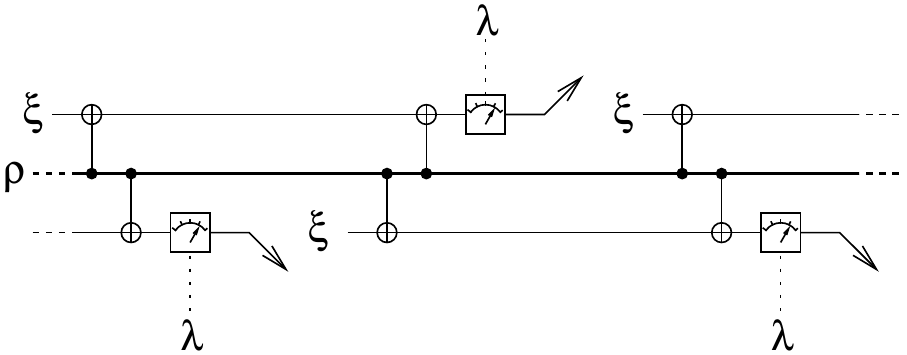}    
    \label{f_NMC_qub}}
\subfigure[Equivalent Markov chain]{
    \includegraphics[scale=1.0]{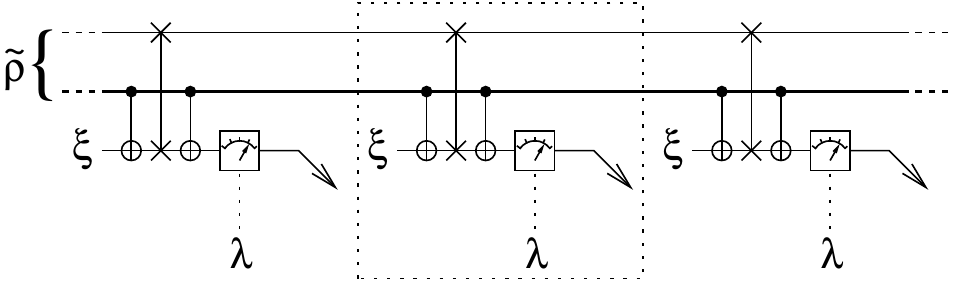}
    \label{f_equivMC_qub}}
\caption{Equivalent one-qubit NM and two-qubit Markov chains. 
\subref{f_NMC_qub} NM chain: the system qubit $\ro$ interacts (collides) twice with the same
reservoir qubits (molecules) $\xi$ via two XOR-gates, with overlap between 'nearest' molecules; 
the post-interaction states of the molecular qubits are measured, yielding the random outcomes $\lambda$.
\subref{f_equivMC_qub} Equivalent Markov chain:
the system-satellite two-qubit 
$\roh$ compound interacts (collides) with the reservoir qubits (molecules) $\xi$ via 
the swap- and two XOR-gates but without the overlap between collisional intervals of different molecules; 
the post-interaction states of the molecular qubits are measured, yielding the random
outcomes $\lambda$.}
\label{f_NMC_MCequiv_qub}
\end{center}
\end{figure*}

To construct the simplest non-Markov chain, let each molecule interact with
the system qubit twice, and let there be an overlap between
subsequent collisional periods, see Fig.~\ref{f_NMC_qub}. We are interested
in stationary chains hence the pattern of double collision will identically
repeat itself along the chain. Note, however, that the starting pattern must
always be a ``broken'' one, i.e., in the beginning there is a molecule that collides
only once with  the central system.

From the classical theory of NM chains with finite memory time
we know that adding suitable amount of memory (bits) to the system yields an
equivalent Markovian chain. Similarly, we attach a memory qubit to the system
qubit. We can make e.g., one distinguished molecule the memory. 
Consider the resulting scheme in Fig.~\ref{f_equivMC_qub}, equivalent with
Fig.~\ref{f_NMC_qub}.
The system+memory compound is a two-qubit composite system whose state 
will be denoted by $\roh_t$ and, as we see, it undergoes independent collisions with 
the rest of the molecules of state $\xi$ each. Our trick is that we repeatedly
swap all system-environment entanglement into the system-memory
compound. 
  
Accordingly, we have obtained a Markovian quantum chain for the
system+memory compound $\roh_t$ instead of the system $\ro_t$ alone. 
Markovian structure (\ref{MC})-(\ref{trUMC}) applies invariably.
The molecule-memory-system compound is a three-qubit system.
The three-qubit unitary operation $U$ is the following:
\begin{equation}\label{U2XOR}
U=\mathrm{XOR}_\mathrm{sys-mol}\,\mathrm{SWAP}_\mathrm{mol-mem}\,\mathrm{XOR}_\mathrm{sys-mol} .
\end{equation}
In the basis $\ket{\mathrm{mol, mem, sys}}=\{\ket{000},\ket{001},\ket{010},\dots\ket{111}\}$, we calculate the
collision matrix (\ref{U2XOR}) times $\exp(i\phi\sigma_y)$, with notations $c=\cos\phi$ and $s=\sin\phi$:
\begin{equation}
U\exp(i\phi\sigma_y)=
\begin{pmatrix}
 c& & & &-s& & & \\
  & & &s& & & &c\\
 s& & & &c& & & \\
  & & &c& & & &-s\\
  & &c& & & &-s& \\
  &s& & & &c& & \\
  & &s& & & &c& \\
  &c& & & &-s& &
\end{pmatrix} .
\end{equation}
According to (\ref{Ml_def}), the two Kraus matrices $M_0$ and $M_1$ 
in the basis $\ket{\mathrm{mem, sys}}=\{\ket{00},\ket{01},\ket{10},\ket{11}\}$
are given by the upper-left and lower-left $4\times4$ blocks, respectively:
\begin{equation}
M_0=
\begin{pmatrix}
 c& &0 & \\
  &0 & &s\\
 s& &0 &  \\
  &0 & &c
\end{pmatrix}
\hspace{3em}
M_1=
\begin{pmatrix}
 0 & &c& \\
  &s& &0 \\
 0 & &s& \\
  &c& &0
\end{pmatrix} .
\end{equation}
Using these Kraus matrices, the non-selective Markov chain (\ref{MC}) reads:
\begin{equation}\label{roh_2XOR}
\roh_{t+1}=
\begin{pmatrix}
 c^2\R{00}{22} & cs \R{03}{21} & cs \R{00}{22} & c^2\R{03}{21} \\
 cs \R{30}{12} & s^2 \R{11}{33} & s^2\R{30}{12} & cs \R{11}{33} \\
 cs \R{00}{22} & s^2 \R{03}{21} & s^2\R{00}{22} & cs \R{03}{21} \\
 c^2\R{30}{12} & cs  \R{11}{33} & cs \R{30}{12} & c^2\R{11}{33}
\end{pmatrix} .
\end{equation}
On the r.h.s. we can identify the diagonal part of the system density matrix
$\ro^t_{00}=\R{00}{22}$ and $\ro^t_{11}=\R{00}{33}$ as well as the specific
correlation $C^t_{x-}=\Tr(\sigma_x\otimes\sigma_-\roh_t)=\R{03}{21}$ between the memory 
and the system.
Observe that all these quantities are invariants along our NM chain. 
Not surprisingly then, the chain will immediately reach the
stationary state $\roh_1=\roh_2\dots=\roh_\infty$ fully parametrized by
the above invariants:
\begin{equation}
\roh_\infty=
\begin{pmatrix}
 c^2\ro^0_{00} & cs C^0_{x-}   & cs \ro^0_{00}  & c^2 C^0_{x-}  \\
 cs C^0_{x+}   & s^2\ro^0_{11} & s^2 C^0_{x+}   & cs\ro^0_{11} \\
 cs \ro^0_{00} & s^2 C^0_{x-}  & s^2\ro^0_{00} & cs C^0_{x-} \\
 c^2 C^0_{x+}  & cs \ro^0_{11} & cs C^0_{x+}    & c^2\ro^0_{11}
\end{pmatrix} .
\end{equation}
The free parameter $C^0_{x\pm}$ depends on the starting ``broken'' pattern.
In the case shown on Figs.~\ref{f_NMC_qub} and \ref{f_equivMC_qub}, 
it can be set to zero, assuming uncorrelated initial satellite memory with
$\langle\sigma_{x}\rangle =0$.
The above stationary state of the memory+system compound then becomes a separable,
disentangled state:
\begin{equation}\label{roh_inf_2XOR}
\roh_\infty=\ro^0_{00}\ket{\Psi_{\phi}}\bra{\Psi_{\phi}}\otimes\ket{0}\bra{0}
+\ro^0_{11}\ket{\Psi'_{\phi}}\bra{\Psi'_{\phi}}\otimes\ket{1}\bra{1} ,
\end{equation}
where the memory state $\vert\Psi'_{\phi}\rangle=s\ket{0}+c\ket{1}$ is not
orthogonal to $\ket{\Psi_{\phi}}=c\ket{0}+s\ket{1}$ (unless $\phi=\pi/4$).
The stationary density matrix $\ro_1=\ro_2\dots=\ro_\infty$ 
of our central system is trivial:
\begin{equation}
\ro_\infty= \Tr_{\mathrm{mem}} \roh_{\infty} =
\begin{pmatrix}
\ro^0_{00}          & 0\\
  0 & \ro^0_{11}
\end{pmatrix}.
\end{equation}
It is the mixture of the basis states with the initial population, as if
$\ro_0$ were ideally measured by the collective of the molecules, like in
case of the Markovian chain, cf. Sec.~\ref{One-qubit_Markov_chain}. 

The repeated-XOR model has trivial invariants, this is the reason it reaches
the asymptotic state in one step already.

\subsection{Distributed XORs with time overlap}\label{Distributed_XORs}
In order to make a NM chain evolve asymptotically towards the stationary
state we must assure that there are fewer trivial invariants. Let us go back 
for a moment to the Markovian chain with single XOR collisions (\ref{UXOR}) and
introduce a NM variant. Suppose that each XOR-collision takes a finite time and
subsequent XOR operations overlap in time. To construct a simplest 
discrete model of time-distributed XOR, we make XOR in two steps separated by unit time.
Let both steps correspond to the following square-root of XOR:
\begin{equation}\label{UsqrtXOR}
\sqrt{\mathrm{XOR}} =
\begin{pmatrix} 
  1 &   &   &  \\ 
    & 1 &   &  \\
    &   & \sqrt{i/2}  & -i\sqrt{i/2} \\ 
    &   & -i\sqrt{i/2} & \sqrt{i/2}  \\ 
\end{pmatrix} .
\end{equation}
(Here the convention $\sqrt{i/2}=e^{i\pi/8}/\sqrt{2}$ has been chosen.)
Our non-Markov model is shown in Fig.~\ref{f_NMC_qub}, with
the XOR collisions replaced by the above $\sqrt{\mathrm{XOR}}$. 

Obviously, we have the
equivalent Markov chain, like in Fig.~\ref{f_equivMC_qub}, and we can perform the
same calculations as before. We calculate the unitary matrix 
\begin{equation}\label{U2sqXOR}
U=\sqrt{\mathrm{XOR}}_\mathrm{sys-mol}\ \mathrm{SWAP}_\mathrm{mol-mem}\,\sqrt{\mathrm{XOR}}_\mathrm{sys-mol}
\end{equation}
times $\exp(i\phi\sigma_y)$, and read out the two Kraus matrices:
\begin{equation}
M_0=
\begin{pmatrix}
 c& &0 & \\
  &i\betac & & \betac \\
  s& &0 &  \\
  & \beta & & -i\beta
\end{pmatrix}
\hspace{3em}
M_1=
\begin{pmatrix}
 0 & &c& \\
  & \betac & & i\betac \\
 0 & &s& \\
  & -i\beta & & \beta
\end{pmatrix} ,
\end{equation}
with $\beta=e^{i\phi}/2$. These Kraus matrices in (\ref{Kraus}) yield, after
direct calculations, the following Markovian chain (\ref{MC}) for the 
memory-system compound:
\begin{equation}\label{roh_2sqXOR}
\roh_{t+1}\!=
\begin{pmatrix}
 c^2 (\roh^t_{00}\!+\!\roh^t_{22}) &
 c \beta \Delta_t &
 c s (\roh^t_{00}\!+\!\roh^t_{22})  &
 i c \betac \Delta_t \\     
 c \betac \Deltac_t &
 1/2(\roh^t_{11}\!+\!\roh^t_{33}) &
 s \betac \Deltac_t &
 2i\betac^2(\roh^t_{11}\!+\!\roh^t_{33}) \\
 c s (\roh^t_{00}\!+\!\roh^t_{22}) &
 s \beta \Delta_t &
 s^2 (\roh^t_{00}\!+\!\roh^t_{22}) &
 i s \betac \Delta_t \\
 -c \beta i\Deltac_t &
 -2i\beta^2(\roh^t_{11}\!+\!\roh^t_{33}) &
 - s \beta i\Deltac_t &
 1/2(\roh^t_{11}\!+\!\roh^t_{33}) \\
\end{pmatrix} ,
\end{equation}
where
\begin{equation}
\Delta_t = -i(\roh^t_{01}+\roh^t_{23})+(\roh^t_{03}+\roh^t_{21})
\end{equation}
and it satisfies a closed evolution equation:
\begin{equation}\label{Deltat}
\Delta_{t+1} = \sin(2\phi)\Delta_t ,
\end{equation}
therefore this is a convenient parametrization.  
It follows from (\ref{roh_2sqXOR}) that the matrix elements not containing $\Delta$
will take their final stationary values immediately after the first step,
just like in our previous model in Sec.~\ref{Repeated_XORs}. However, the elements with  
$\Delta$ show an exponential relaxation (\ref{Deltat}) toward zero 
if $\sin(2\phi)\neq0$. The relaxation of $\Delta_t$ governs the
asymptotic diagonalization of the system density matrix:
\begin{equation}
\ro_t = \Tr_{\mathrm{mem}} \roh =
\begin{pmatrix}
 \ro^{t=0}_{00} & (1+i\sin(2\phi))\Delta_t/2 \\
 (1-i\sin(2\phi))\Deltac_t/2 &\ro^{t=0}_{11}
\end{pmatrix} .
\end{equation}
Indeed, for $t=\infty$ one has $\Delta_\infty=0$, the system density matrix
$\ro_\infty$ becomes diagonal with the initial populations $\ro^0_{00},\ro^0_{11}$.
As to the stationary density matrix $\roh_\infty$ of the memory-system compound,  
the Markov chain (\ref{roh_2sqXOR}) yields a separable state again,   
like in our previous model in Sec.~\ref{Repeated_XORs}. This time we get
\begin{equation}\label{roh_inf_2sqXOR}
\roh_\infty=\ro^0_{00}\ket{\Psi_{\phi}}\bra{\Psi_{\phi}}\otimes\ket{0}\bra{0}
            +\ro^0_{11}\ket{\Psi'_{\phi}}\bra{\Psi'_{\phi}}\otimes\ket{1}\bra{1} ,
\end{equation}
where the memory state $\vert\Psi'_{\phi}\rangle=(\ket{0}+i e^{2i\phi}\ket{1})/\sqrt2$,
different from $\vert\Psi'_{\phi}\rangle$ in the model in Sec.~\ref{Repeated_XORs}, is
never orthogonal to $\ket{\Psi_{\phi}}$. This is what makes our system \emph{quantum} NM, a distinction discussed in the next section.

\section{Classical/Quantum non-Markovianity}
\label{Classical_Quantum_non-Markovianity}
Our construction of NM quantum chains may become universal if we allow for
more than two repeated collisions with a single molecule and/or overlaps of
collision periods between more than two molecules. Whether or not an arbitrary
NM chain will be reducible to ours remains an open theoretical issue. Our
class of NM chains is unique for at least one thing: we can always
identify the satellite memory to make the time evolution of the resulting system-memory 
compound a Markovian chain. We can always
read-out from the circuit of the NM chain how many qubits are needed for the satellite
memory. This number of qubits is a useful \emph{count} of 
non-Markovianity of our chain, e.g., in computational simulation this number
gives an upper bound on how much data should be stored dynamically together with the data of our 
system of interest.

However, this count may be significantly larger than 
the amount of information to be contained in the satellite.  
Suppose that we have determined the minimum number of qubits needed for the satellite, 
so the count of non-Markovianity is known. Then a suitable informatic quantity, like the mutual 
information \cite{NieChu00,Dio11} of the satellite memory (M) and the system (S) might 
play the role of non-Markovianity \emph{measure} $\mu_{NM}$. In the stationary regime one gets
\begin{equation}
\mu_{NM}=I(S:M)=H(S)+H(M)-H(S,M)
\end{equation}
where $H$ stands in turn for the von Neumann entropy of the S-state $\Tr_{\mathrm{mem}}\roh_\infty=\ro_\infty$, 
of the M-state $\Tr_{\mathrm{sys}}\roh_\infty$, and of the composite state $\roh_\infty$.

We should stop for a second, and distinguish \emph{quantum} from \emph{classical} non-Markovianity.
The attentive reader may have noticed that our construction of quantum NM chains 
in itself has nothing particular for quantum chains, it roots in a similar construction
of classical NM chains. In fact, any classical NM chain can be represented by an equivalent
Markovian chain if we assign a sufficient satellite memory. The minimum size of the satellite memory
(e.g.: in bits) is the natural count of non-Markovianity,
the mutual information may be the measure of non-Markovianity.  
In the case of a generic quantum NM chain, it may happen that the minimum satellite memory still 
consists of bits, qubits are not required at all. In this case we say that the chain is 
\emph{classically NM}, and its quantum non-Markovianity is zero. It is remarkable that both NM
chains in Sec.~\ref{One-qubit_non-Markov_Chains} have turned out to be quantum NM. If we
look at the composite states (\ref{roh_inf_2XOR},\ref{roh_inf_2sqXOR}) of the system+satellite 
compound, we see that classical satellite bits would not work, we need satellite qubits. 
Although we got zero stationary entanglement between the system and satellite, 
it does not mean the lack of quantumness. The price of getting 
rid of quantum non-Markovianity cannot be paid in classical bit instead of qubit.  
The count of quantum non-Markovianity is 1 for both NM chains.
In order to distinguish quantum from classical non-Markovianity
we use the notion of discord \cite{OllZur02}.
The classical measure of non-Markovianity may be defined by
\begin{equation}
\mu_{NM}^{cl}=J(S:M)_{\{\Pi^M_j\}}=H(S)-H(S\vert\{\Pi^M_j\})
\end{equation}
where the rightmost term means the average von Neumann entropy of S when the satellite
memory is undergoing the projective measurement via the set $\Pi^M_j$.  
The quantum non-Markovianity measure is the discord itself:
\begin{equation}
\mu_{NM}^{qu}=I(S:M)-J(S:M)_{\{\Pi^M_j\}}.
\end{equation}
The sum of quantum and classical non-Markovianity measures yields the total measure $\mu_{NM}$.

We can check our proposal on the two models of Sec.~\ref{One-qubit_non-Markov_Chains}.
Since the memory states $\ket{\Psi_\phi},\ket{\Psi_\phi'}$ are non-orthogonal, we get non-zero
discord. If they were orthogonal, we could get zero discord, and also we could measure the
memory after each collision so that a single classical bit ${0,1}$ could be retained instead
of the memory qubit: the chain would be classically NM in the stationary regime, with non-Markovianity
count 1 (bit).  

\section{Brief summary and outlook}

In the framework of the abstract collision model of system-reservoir interactions,
we constructed quantum chains meeting the minimal requirements for NM stationary time evolution.
A method of systematic construction of the equivalent Markovian quantum chains 
using explicit memory allocation is discussed. Due to the transparent NM structure,
we can always identify a well-defined part of the reservoir as the memory, this part is called
the satellite memory of the system. The time evolution of the system-plus-satellite-memory
compound is Markovian. We suggest a novel distinction: a given quantum NM chain is either
quantum NM (if the satellite requires qubits) or classical NM (if bits suffice for the
satellite). Accordingly, we suggest the  numbers of satellite qubits and
bits, respectively, as counts of quantum/classical non-Markovianity of a given quantum
chain. The mutual information and discord is proposed to measure non-Markovianity and
quantum non-Markovianity, respectively. 

Two examples of one-qubit NM chains are discussed. 
In both examples, the corresponding satellite memory is a single
qubit. The stationary state is exactly calculable, the non-vanishing discord of the
satellite qubit w.r.t. the system qubit indicates that our examples are quantum non-Markovian.  

Although our calculations are restricted for non-selective NM chains, the structural transparency 
allows for the selective refinements. Option of monitoring the NM system is inherent
in the model.  
  
\begin{figure}
\centering
\includegraphics[scale=1.0]{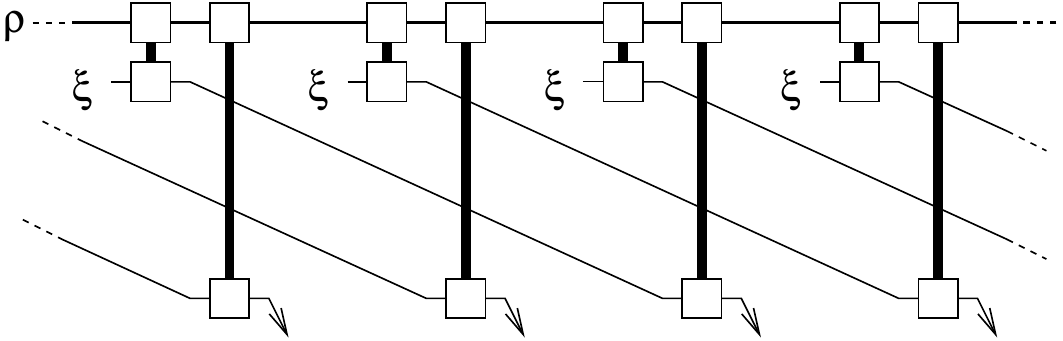}
\caption{NM quantum chain of advanced overlap: the system interacts with each independent 
molecule twice, with overlap between collisional intervals of 'nearest' and 'next-to-nearest' molecules.
(Boxes connected with vertical lines represent the bipartite unitary collisions.) 
}
\label{f_Outlook}
\end{figure}

For achieving a higher level non-Markovianity than discussed in our work (Fig.~\ref{fa_diosi}), 
one can construct a similar chain with more collisions per molecule and/or longer intervals of overlap. 
In Fig.~\ref{f_Outlook}, e.g., the number of collisions is invariably two whereas the lengths of
overlapping intervals have been increased for three collisions. In return, the equivalent Markovian structure 
needs two satellite qubits instead of one, the non-Markovianity count is 2.
The ability to handle complex NM processes may pave the way to 
the construction of environmental interaction models which better approximate reality.

Support by the Hungarian Scientific Research Fund under Grant No. 75129
and support by the EU COST Action MP100 are acknowledged. The authors thank Tam\'as Geszti
for useful discussions.

\end{document}